\documentclass{emulateapj}
\usepackage{apjfonts}
\lefthead{HAN \& AN} \righthead{PLANETARY CAUSTICS}

\newcommand{\svec}{\bold s}
\newcommand{\rvec}{\bold r}

\def\eqalign#1{\null\,\vcenter{\openup\jot
        \ialign{\strut\hfil$\displaystyle{##}$&$
        \displaystyle{{}##}$\hfil \crcr#1\crcr}}\,}

\begin{document}
\title{Properties of Planetary Caustics in Gravitational Microlensing}

\author{Cheongho Han}
\affil{Department of Physics, Institute for Basic Science
Research, Chungbuk National University, Chongju 361-763, Korea;\\
cheongho@astroph.chungbuk.ac.kr}



\begin{abstract}
Although some of the properties of the caustics in planetary microlensing 
have been known, our understanding of them is mostly from scattered 
information based on numerical approaches.  In this paper, we conduct 
a comprehensive and analytic analysis of the properties of the planetary 
caustics, which are one of the two sets of caustics in planetary microlensing, 
those located away from the central star.  Under the perturbative 
approximation, we derive analytic expressions for the location, size, and 
shape of the planetary caustic as a function of the star-planet separation 
and the planet/star mass ratio.  Based on these expressions combined with 
those for the central caustic, which is the other set of caustics located 
close to the central star, we compare the similarities and differences 
between the planetary and central caustics.  We also present the expressions 
for the size ratio between the two types of caustics and for the condition 
of the merging of the two types of caustics.  These analytic expressions 
will be useful in understanding the dependence of the planetary lensing 
behavior on the planet parameters and thus in interpreting the planetary 
lensing signals.
\end{abstract}


\section{Introduction}

Microlensing is one of the most powerful methods that can be used to 
search for extrasolar planets \citep{mao91, gould92}.  Recently, two 
robust microlensing detections of  exoplanets were reported by 
\citet{bond04} and \citet{udalski05}.  

The signal of a planetary companion to microlens stars is a short-duration 
perturbation to the smooth standard light curve of the primary-induced 
lensing event occurring on a background source star.  The planetary 
perturbation occurs when the source star passes close to the caustic.  
The caustic represents the set of source positions at which the 
magnification of a point source becomes infinite.  Studies of the 
properties of the caustic are important because the characteristics of 
the planetary perturbations in the microlensing light curve depend 
critically on the properties of the caustic.  For example, the location 
of the perturbation on the lensing light curve depends on the location 
of the caustic.  The duration of the perturbation and the probability of 
detecting the perturbation are proportional to the caustic size.  In 
addition, the pattern of the perturbation is closely related to the 
caustic shape.  Therefore, it is essential to understand the properties 
of caustics for the interpretation of the planetary lensing signals.

Although some of the properties of the caustics in planetary microlensing 
have been known, our knowledges of them are mostly from scattered 
information based on numerical approaches.  The problem of the numerical 
approach is that the dependence of the planetary lensing behavior on the 
planet parameters of the star-planet separation $s$ (normalized by the 
Einstein ring radius $\theta_{\rm E}$) and the planet/star mass ratio 
$q$ is not clear.  There have been several attempts to overcome this 
ambiguity using analytic methods.  By treating the planet-induced deviation 
as a perturbation, \citet{dominik99} and \citet{an05} derived analytic 
expressions for the locations of the {\it central} caustic, which is one 
of the two sets of caustics of the star-planet lens system located close to 
the primary star.  Based on a similar perturbative approach, \citet{asada02} 
provides analytic expressions for the locations of the lensing images.  
\citet{bozza00} derived analytic expressions for the locations of not only 
the central caustic but also the {\it planetary} caustic, the other set of 
caustics, which are located away from the central star.  However, there has
been no analytic work on the detailed properties of the caustics such as 
the location, size, and shape, except the very recent work of \citet{chung05} 
(hereafter paper I) on the central caustics.

Following paper I, we conduct a comprehensive and analytic analysis on the 
properties of the planetary caustics.  Under the perturbative approximation, 
we derive analytic expressions for the location, size, and shape of the 
planetary caustics as a function of $s$ and $q$.  Based on these expressions 
combined with those for the central caustics derived in paper I, we compare 
the similarities and differences between the planetary and central caustics. 
We provide an expression for the size ratio between the two types of 
caustics.  We also derive an expression for the condition of the merging 
of the two types of caustics.  We finally discuss the validity of the 
perturbative approximation.

\section{Empirical Known Properties}

A planetary lensing is described by the formalism of a binary lens with a 
very low-mass companion.  Because of the very small mass ratio, planetary 
lensing behavior is well described by that of a single lens of the primary 
star for most of the event duration.  However, a short-duration perturbation 
can occur when the source star passes the region around the caustics, which 
are important features of binary lensing.

The caustics of binary lensing form a single or multiple closed figures
where each of which is composed of concave curves (fold caustics) that
meet at cusps.  For a planetary case, there exist two sets of disconnected 
caustics.  One `central caustic' is located close to the host star.  The 
other `planetary caustic' is located away from the host star and its number
is one or two depending on whether the planet lies outside ($s>1$) or inside 
($s<1$) the Einstein ring.  The size of the caustic, which is directly 
proportional to the planet detection efficiency, is maximized when the planet 
is located in the `lensing zone', which represents the range of the star-planet
separation of $0.6\lesssim s\lesssim 1.6$.  The planetary caustic is always 
bigger than the central caustic.

\section{Analytic Approach}

\begin{figure}[t]
\epsscale{1.2}
\plotone{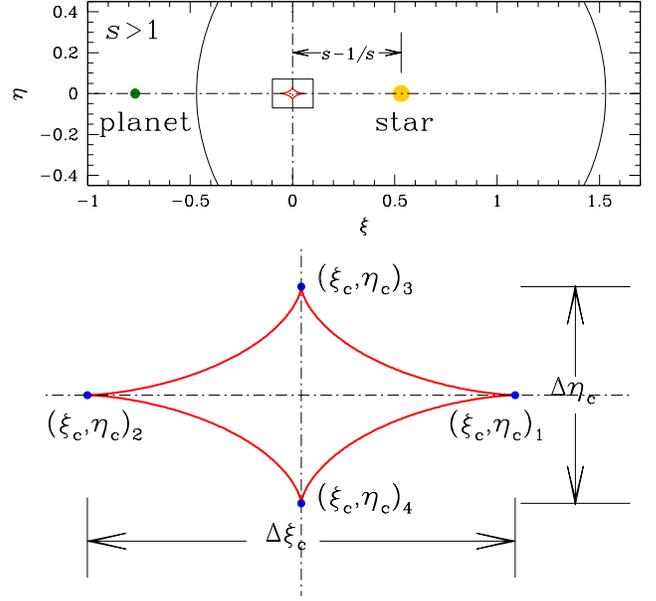}
\caption{\label{fig:one}
The location and shape of the planetary caustic of a planetary system with 
a star-planet separation greater than the Einstein ring radius ($s>1$).  
The upper panel shows the location of the star, the planet, and the 
resulting location of the planetary caustic.  The coordinates are centered 
at the center of the planetary caustic, which is located on the star-planet 
axis with a separation vector $\rvec=\svec(1-1/s^2)$ from the position of 
the star, where $\svec$ is the position vector of the planet from the host 
star.  The circle centered at the position of the star is the Einstein ring.
The lower panel shows a blow-up of the region around the caustic enclosed 
by a box.  Also marked are the definitions of the horizontal ($\Delta\xi_c$) 
and vertical ($\Delta\eta_c$) widths of the caustic and the designations of 
the individual cusps of the caustic.
}\end{figure}

\begin{figure}[t]
\epsscale{1.1}
\plotone{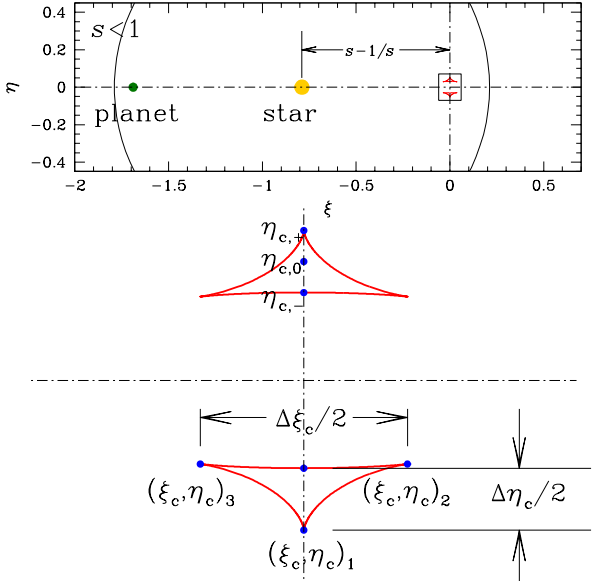}
\caption{\label{fig:two}
The location and shape of the planetary caustic of a planetary system with 
a star-planet separation less than the Einstein ring radius ($s<1$).  
Notations are same as in Fig.~\ref{fig:one}.
}\end{figure}

We start from the formula of \citet{bozza00} for the position of the 
planetary caustics (eqs. [49] and [50] of his paper).  Keeping up to 
the first order term, the formula are expressed as 
\begin{equation}
\xi_c \simeq
q^{1/2} \left( \kappa-{1\over \kappa}+{\kappa\over s^2}\right) \cos\theta,
\label{eq1}
\end{equation}
\begin{equation}
\eta_c \simeq
q^{1/2}\left( \kappa-{1\over \kappa}-{\kappa\over s^2}\right) \sin\theta,
\label{eq2}
\end{equation}
where $\theta$ is a variable and 
\begin{equation}
\kappa(\theta)=\left[ {\cos 2\theta \pm \sqrt{s^4-\sin^2 2 \theta}
\over s^2-1/s^2 }\right]^{1/2}.
\label{eq3}
\end{equation}
In these expressions, the coordinates are centered at the position on the 
star-planet axis with a separation vector from the position of the star of 
\begin{equation}
\rvec=\svec\left(1-{1\over s^2}\right),
\label{eq4}
\end{equation} 
where $\svec$ is the position vector of the planet from the star normalized 
by $\theta_{\rm E}$ (see Figs.~\ref{fig:one} and \ref{fig:two}).  The 
origin of the coordinates corresponds to the center of the planetary caustic.  
For the pair of the planets with separations $s$ and $1/s$, the centers of 
the caustics are separated from the star by the same distance (because 
$|\rvec(s)| = |\rvec(1/s)|$) but directed toward opposite directions (because 
${\rm sign} [\rvec(s)]\neq {\rm sign} [\rvec(1/s)]$).  Therefore, the center 
of the caustic is located on the same and opposite sides of the planet with 
respect to the position of the star for the planets with $s>1$ and $s<1$, 
respectively.  If one defines the lensing zone as the range of the planetary 
separation for which the planetary caustic is located within the Einstein 
ring, the exact range of the lensing zone is 
\begin{equation}
{\sqrt{5}-1\over 2} \leq s \leq {\sqrt{5}+1\over 2}.
\label{eq5}
\end{equation}
To the first-order approximation, the size of the planetary caustic is 
proportional to $q^{1/2}$ as shown in equations~(\ref{eq1}) and (\ref{eq2}).  
We will discuss the deviation of the approximation from the exact value in 
\S\ 4.3.

\subsection{For Planets with $s>1$}

In this case, between the two values of $\kappa$ in equation~(\ref{eq3})
only the one with `+' sign is valid because the other one with `$-$' 
sign results in $\kappa^2<0$.  As a result, there exists only a single 
set of caustics for planets with $s>1$ as shown in Figure~\ref{fig:one}.

The planetary caustic of the planet with $s>1$ is composed of four cusps, 
with two of them are located on the $\xi$ axis and the other two are 
located on the $\eta$ axis (see Fig.~\ref{fig:one}).  The positions of 
the individual cusps, $(\xi_c,\eta_c)_i$, corresponds to the cases of 
$\sin\theta=0$ (for the two cusps on the $\xi$ axis) and $\cos\theta=0$ 
(for the other two cusps on the $\eta$ axis).  Then, the positions of the 
cusps on the $\xi$ and $\eta$ axes are expressed respectively as
\begin{equation}
(\xi_c,\eta_c)_{1,2} \simeq 
\left( 
{2q^{1/2} \over s\sqrt{s^2-1}},0
\right),
\label{eq6}
\end{equation}
\begin{equation}
(\xi_c,\eta_c)_{3,4} \simeq 
\left( 0,{2q^{1/2} \over s\sqrt{s^2+1}}
0, 
\right).
\label{eq7}
\end{equation}

If we define the horizontal and vertical widths of the planetary caustic
as the separations between the cusps on the individual axes (see 
Fig.~\ref{fig:one}), the widths are expressed respectively as
\begin{equation}
\Delta \xi_c \simeq
{4q^{1/2} \over s\sqrt{s^2-1}} \rightarrow 
{4q^{1/2}\over s^2}\left( 1+ {1\over 2s^2}\right),
\label{eq8}
\end{equation}
\begin{equation}
\Delta \eta_c \simeq
{4q^{1/2} \over s\sqrt{s^2+1}} \rightarrow 
{4q^{1/2}\over s^2}\left( 1- {1\over 2s^2}\right),
\label{eq9}
\end{equation}
where the expressions after the arrow are those evaluated to the first
non-vanishing order in $s$ in the limiting case of $s\gg 1$.  Then, the 
vertical/horizontal width ratio is expressed as
\begin{equation}
{\cal R}_c =
{\Delta\xi_c \over \Delta \eta_c} \simeq
\left( {1-1/s^2 \over 1+1/s^2}  \right)^{1/2} \rightarrow
1-{1\over s^2}.
\label{eq10}
\end{equation}
In the limiting case of $s\gg 1$, $\Delta\xi\sim \Delta\eta\propto s^{-2}$
and ${\cal R}_c\sim 1$, i.e.  the caustic size decreases as $s^{-2}$ and 
the shape becomes less elongated as the star-planet separation increases.

\subsection{For Planets with $s<1$}

In this case, $\kappa$ in equation~(\ref{eq3}) is valid only in the 
following range of $\theta$
\begin{equation}
\theta\pm {\pi\over 2} < {1\over 2} \sin^{-1} s^2.
\label{eq11}
\end{equation}
For $\theta$ within these ranges, there are two possible values of $\kappa$ 
corresponding to the signs.  As a result, there exist two sets of caustics 
for planets with $s<1$; one above and the other below the star-planet 
axis (see Fig.~\ref{fig:two}).

Each of the caustics for the planet with $s<1$ is composed of three cusps.
One of them is located on the $\eta$ axis but the other two are not located 
on either of the axes.  The caustic meets the $\eta$ axis at $\eta_{c,+}=
2q^{1/2}/[s (1+s^2)^{1/2}]$ and $\eta_{c,-}=2q^{1/2} (1-s^2)^{1/2}/s$ when 
$\cos\theta=0$ (see Fig.~\ref{fig:two}).  Among these two positions, the 
former corresponds to the cusp, and thus the location of the on-axis cusp is
\begin{equation}
(\xi_{c},\eta_{c})_1 \simeq
\left(0, \pm{2q^{1/2}\over s\sqrt{1+s^2}} \right),
\label{eq12}
\end{equation}
where the sign `$\pm$' is for the cusps located above and below the 
star-planet axis, respectively.

If we define the vertical width of the caustic as the separation between 
the two crossing points at $\eta_{c,+}$ and $\eta_{c,-}$, the width is 
expressed as
\begin{equation}
{\Delta\eta_c\over 2} 
\simeq {2q^{1/2}\over s}
\left( {1\over \sqrt{1+s^2}}-\sqrt{1-s^2}\right) \rightarrow
q^{1/2} s^3, 
\label{eq13}
\end{equation}
where the factor `1/2' is included into consideration that there exist 
two planetary caustics for planets with $s<1$ and the expression after 
the arrow is that evaluated to the first non-vanishing order in $s$ in 
the case of $s\ll 1$.  By defining the center of {\it each} caustic as 
the midpoint between the two crossing points (see Fig.~\ref{fig:two}), 
its position is expressed as 
\begin{equation}
\eta_{c,0} 
\simeq \pm {q^{1/2}\over s} 
\left( {1\over \sqrt{1+s^2}}+\sqrt{1-s^2}\right) \rightarrow
{2q^{1/2}\over s} \left( 1-{1\over 2}s^2\right).
\label{eq14}
\end{equation}

The other two cusps occurs when $d\xi/d\theta=0$ (or $d\eta/d\theta=0$). 
This condition is satisfied when $\cos^2 2\theta=1-3s^4/4$ (or $\sin^2 
2\theta=3s^4/4$).  Then, combined with the possible range in 
equation~(\ref{eq11}), the values of $\theta$ corresponding to the off-axis 
cusps are found to be
\begin{equation}
\theta_0={\pi\over 2}\pm{1\over 2}\sin^{-1}\left({\sqrt{3}\over 2}s^2\right).
\label{eq15}
\end{equation}
With this value combined with equations~(\ref{eq1}) and (\ref{eq2}), the 
positions of the off-axis cusps are expressed as
\begin{equation}
\eqalign{
(\xi_c,\eta_c)_{2,3} \simeq
[ \pm q^{1/2}\left(\kappa_0-{1/\kappa_0}+
{\kappa_0/s^2}\right)\cos\theta_0, \cr 
\pm q^{1/2}\left(\kappa_0-{1/\kappa_0}-
{\kappa_0/s^2}\right)\sin\theta_0 ],\cr} 
\label{eq16}
\end{equation}
where $\kappa_0=\kappa(\theta_0)$.  In the limiting case of $s\ll 1$, 
equation~(\ref{eq16}) is approximated as
\begin{equation}
(\xi_c,\eta_c)_{2,3} \rightarrow
\left( \pm{3\sqrt{3}q^{1/2}s^3\over 8}, \pm {{2q^{1/2}\over s}}\right),
\label{eq17}
\end{equation} 
because $\kappa_0\rightarrow s(1+s^2/4)$, $\sin\theta_0\rightarrow 1$, and 
$\cos\theta_0\rightarrow \sqrt{3}s^2/4$.  By defining the horizontal width 
as the separation between the two off-axis cusps, the width is expressed as 
\begin{equation}
{\Delta\xi_c\over 2} \simeq
2q^{1/2} \left(\kappa_0-{1\over\kappa_0}+
{\kappa_0\over s^2}\right)\cos\theta_0 
\rightarrow {3\sqrt{3}\over 4} q^{1/2}s^3.
\label{eq18}
\end{equation}
Once again, the factor `1/2' is included into consideration that there are 
two planetary caustics.  From equations~(\ref{eq13}) and (\ref{eq18}), the 
vertical/horizontal width ratio is expressed as
\begin{equation}
{\cal R}_c \simeq
{ (1+s^2)^{-1/2} - (1-s^2)^{1/2}
\over
s(\kappa_0-{1/\kappa_0}+{\kappa_0/s^2}) \cos\theta_0}
\rightarrow {4\over 3\sqrt{3}} \left( 1-{5\over 12} s^2\right).
\label{eq19}
\end{equation}
In the limiting case of $s\ll 1$, each caustic shrinks as $\propto s^{3}$, 
c.f. $\propto s^{-2}$ for planets with $s>1$, and 
${\cal R}_c\sim 4/3\sqrt{3}\sim 0.770$, c.f.\ ${\cal R}_c\sim 1$ for planets 
with $s>1$.

\begin{figure}[t]
\epsscale{1.2}
\plotone{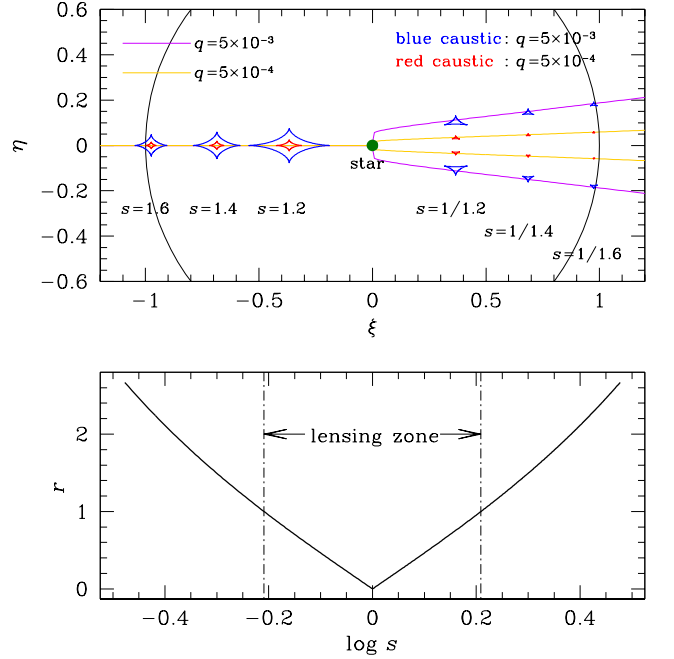}
\caption{\label{fig:three}
Upper panel: example planetary caustics of several planetary systems with 
  different values of the star-planet separation $s$ and the planet/star 
  mass ratio $q$.  The orange and purple curves represent the loci of the 
  center of the caustic as a function of $s$ for planets with 
  $q=5\times 10^{-3}$ and $q=5\times 10^{-4}$, respectively.  The 
  coordinates are centered at the position of the planet-hosting star.
  The circle is the Einstein ring.
Lower panel: separation between the caustic center and the position of the 
  planet-hosting star as a function of $s$.
}
\end{figure}

\section{Caustic Properties in Planetary Lensing}

Based on the analytic expressions derived in the previous section, we 
now investigate how the properties of the planetary caustics such as 
the location, size, and shape vary depending on $s$ and $q$.  We also 
compare the properties of the planetary caustics with those of the central 
caustics.

\subsection{Properties of Planetary Caustics}

\begin{figure}[t]
\epsscale{1.2}
\plotone{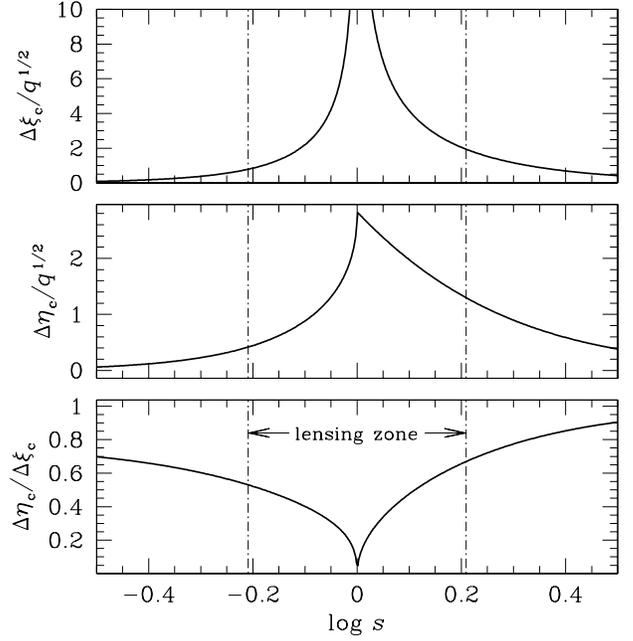}
\caption{\label{fig:four}
Variation of the size (normalized by $q^{1/2}$) and shape (as measured 
by the vertical/horizontal width ratio) of the planetary caustic as a 
function of the star-planet separation $s$.
}\end{figure}

In the upper panel of Figure~\ref{fig:three}, we present example planetary
caustics of several planetary systems with different $s$ and $q$.  In the 
lower panel, we present the separation of the caustic from the planet-hosting 
star as a function of $s$.  In Figure~\ref{fig:four}, we also present the 
variation of the caustic size (as measured by the horizontal and vertical 
widths) and the shape (as measured by the vertical/horizontal width ratio) 
as a function of $s$.

The properties of the planetary caustics found from the figures and the 
dependence of these properties on the planet parameters are as follows.
\begin{enumerate}
\item
For $s>1$, the location of the caustic center depends on $s$ but not on  
$q$.  On the other hand, for planets with $s<1$, the caustic location depends 
on both $s$ and $q$.  In this case, the caustic is located farther away from 
the star-planet axis as $q$ increases (see eq.~[\ref{eq14}]).
\item
Although the caustic size depends on the mass ratio as $\propto q^{1/2}$,
the shape of the caustic does not depend on $q$ and solely dependent on $s$
(see eqs.~[\ref{eq10}] and [\ref{eq19}]).
\item
The rate of decrease of the caustic size with the increase of $|\log s|$ 
are different for planets with $s>1$ and $s<1$.  Compared to the caustic 
of the planet with $s>1$, the rate of decrease is steeper for the planet 
with $s<1$.  In the limiting cases of $s\gg 1$ and $s\ll 1$, the caustic 
sizes decrease as $\propto s^{-2}$ and $\propto s^3$ for planets with $s>1$ 
and $s<1$, respectively (see eqs.~[\ref{eq8}], [\ref{eq9}], [\ref{eq13}], 
and [\ref{eq18}]).
\end{enumerate}

\subsection{Comparison with Central Caustics}

\citet{chung05} presented the analytic expressions for the location, cusp 
positions, width, and shape of the central caustics analogous to those 
presented in the previous section for the planetary caustics.  The 
expressions for the location of the central caustic, analogous to 
equations~(\ref{eq1}) and (\ref{eq2}) for the planetary caustic, are
\begin{equation}
\xi_c \simeq  q {s+1/s+2(\cos^3\phi-2\cos\phi) \over 
(s+1/s-2\cos\phi)^2 },
\label{eq20}
\end{equation}
\begin{equation}
\eta_c \simeq -q{2\sin^3\phi \over   (s+1/s-2\cos\phi)^2 },
\label{eq21}
\end{equation}
where $\phi$ is a variable and the coordinates are centered at the position 
of the host star.  There exists a single central caustic regardless of $s$ 
and it has an elongated asteroid shape with four cusps, of which two are 
located on the $\xi$ axis and the other two are off the axis.  The analytic 
expressions for the positions of the individual cusps, which are analogous 
to equations~(\ref{eq6}) and (\ref{eq7}) for the planetary caustic with 
$s>1$ and to equations~(\ref{eq12}) and (\ref{eq16}) for the planetary 
caustic with $s<1$, are
\begin{equation}
(\xi_c,\eta_c)_{1,2} \sim \left[
 \pm {q\over (1\pm s)(1\pm 1/s)}, 0 \right],
\label{eq22}
\end{equation}
\begin{equation}
(\xi_c,\eta_c)_{3,4} \sim
 \left[0, \pm{2q |\sin^3\phi_c|\over (s+1/s-2\cos\phi_c)^2} \right],
\label{eq23}
\end{equation}
where $\cos\phi_c = (3/4)(s+1/s) \{ 1- [1-(32/9) (s+1/s)^{-2}]^{1/2}  \}$.
The horizontal and vertical widths of the central caustic defined as the 
separations between the cusps on and off the star-planet axis are expressed 
respectively as
\begin{equation}
\Delta\xi_c = {4q\over (s-1/s)^{2}},
\label{eq24}
\end{equation}
\begin{equation}
\Delta\eta_c = 
{4q\over (s-1/s)^{2}}
{(s-1/s)^2|\sin^3\phi_c| \over (s+1/s-2\cos\phi_c)^2},
\label{eq25}
\end{equation}
which are analogous to those in equations~(\ref{eq8}) and (\ref{eq9}) 
for the planetary caustic with $s>1$ and to equations~(\ref{eq13}) and 
(\ref{eq18}) for the planetary caustic with $s<1$.  
Then, the width ratio of the central caustic is
\begin{equation}
{\cal R}_c =  {(s-1/s)^2|\sin^3\phi_c| \over (s+1/s-2\cos\phi_c)^2},
\label{eq26}
\end{equation}
which is analogous to those in equations~(\ref{eq10}) and (\ref{eq19}) for 
the planetary caustics with $s>1$ and $s<1$, respectively.  In the limiting 
cases of $s\gg 1$ and $s\ll 1$, the size of the central caustic decreases 
respectively as 
\begin{equation}
\Delta\xi_c \sim \Delta\eta_c \rightarrow 
\cases{
4q/s^2 & for $s\gg 1$,\cr
4qs^2  & for $s\ll 1$.\cr
}
\label{eq27}
\end{equation}

\begin{figure}[thb]
\epsscale{1.2}
\plotone{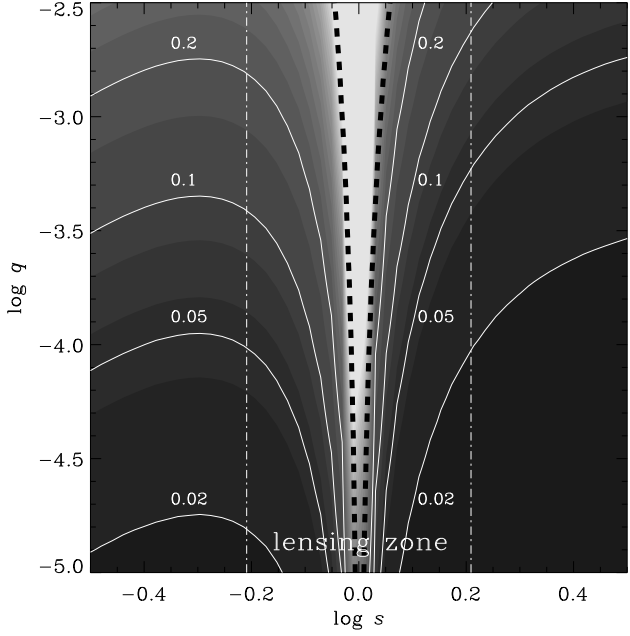}
\caption{\label{fig:five}
The size ratio between the planetary and central caustics as a function 
of the star-planet separation $s$ and planet/star mass ratio $q$.  For a
representative quantity of the caustic size, we use the horizontal width.
The region enclosed by the thick dashed lines represents the area in which 
the planetary and central caustics merge together, resulting in gradual 
obliteration of the distinction between the two types of caustics.
}\end{figure}

The planetary and central caustics have the following similarities and
differences.
\begin{enumerate}
\item
Unlike the planetary caustic, the pair of the central caustics with 
separations $s$ and $1/s$ are identical as demonstrated by the fact that 
the inversion $s \leftrightarrow 1/s$ in equations~(\ref{eq20}) and 
(\ref{eq21}) results in the same expressions.  
\item
While the dependence of the size of the planetary caustic on the planet/star 
mass ratio is $\propto q^{1/2}$, the dependence of the central caustic is
$\propto q$.  Therefore, the planetary caustic shrinks much more slowly with 
the decrease of the planet mass than the central caustic.
\item
For planets with $s>1$, the rate of decrease of the size of the central 
caustic with the increase of $|\log s|$ is similar to that of the planetary 
caustic with $s>1$, i.e.\ $\Delta\xi \propto s^{-2}$ (see eqs.~[\ref{eq8}] and
[\ref{eq27}]), but smaller than that 
of the planetary caustic with $s<1$, which shrinks as $\propto s^{3}$
(see eq.~[\ref{eq18}]). 
\end{enumerate}

\begin{figure*}[t]
\epsscale{0.9}
\plotone{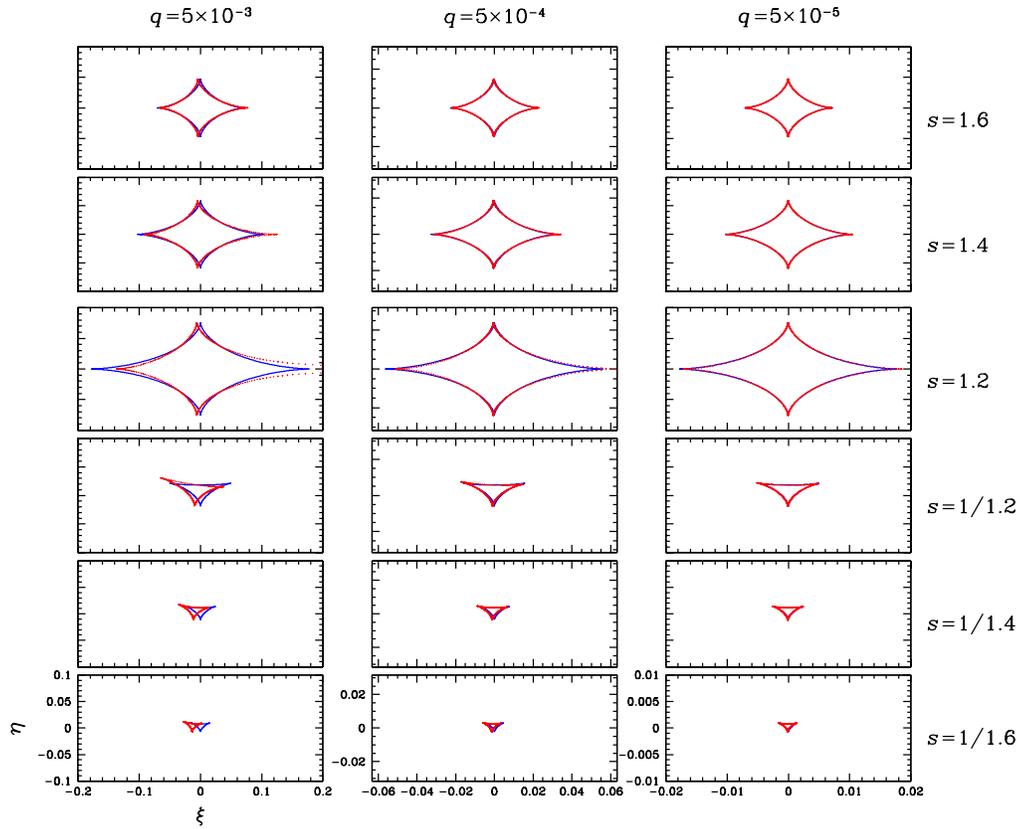}
\caption{\label{fig:six}
Comparison of the planetary caustics based on the analytic (blue caustic) 
numerical (red caustic) computations for various values of the star-planet 
separation $s$ and the planet/star mass ratio $q$.  The coordinates are 
centered at the center of the individual caustics.  The scales of the 
individual panels are set so that the caustics with the same $s$ appear 
to have the same size.
}\end{figure*}

Then, what is the size ratio between the planetary and central caustics.  
If we use the horizontal width as a representative quantity for the caustic 
size, the size ratio between the two types of the caustics is found from 
equations~(\ref{eq8}), (\ref{eq18}), and (\ref{eq24}) and expressed as 
\begin{equation}
{\Delta\xi_{c,{\rm c}}\over \Delta\xi_{c,{\rm p}}} = 
\cases{
q^{1/2} /(1-s^{-2})^{3/2}  & for $s>1$, \cr
q^{1/2} /[(s-s^{-1})^2(\kappa_0-\kappa_0^{-1}+\kappa_0s^{-2}) \cos\theta_0] & for $s<1$, \cr
}
\label{eq28}
\end{equation}
where the additional subscripts `p' and `c' denote the planetary and central
caustics, respectively.  In Figure~\ref{fig:five}, we present the size ratio 
as a function of $s$ and $q$.  Since $\Delta\xi_{c,{\rm c}}\propto q$ while 
$\Delta\xi_{c,{\rm p}}\propto q^{1/2}$, the dependence of the size ratio on 
the mass ratio is $\Delta\xi_{c,{\rm c}}/\Delta\xi_{c,{\rm p}}\propto q^{1/2}$.
For a given mass ratio, the size ratio is maximized at around $s\sim 1$ and 
decreases rapidly with the increase of $|\log s|$.\footnote{For the case of 
$s<1$, the change rate of the size ratio is reversed as $|\log s|$ further 
increases beyond a critical value ($|\log s|\sim -0.3$ or $s\sim 0.5$).
However, this reversal occurs at the separation beyond the lensing zone.
}

As $s\rightarrow 1$,  the location of the planetary caustic, i.e.\ $\rvec
= \svec(1-1/s^2)$, approaches the position of the central star, around which
the central caustic is located.  Then the two types of the caustics eventually 
merge together, resulting in gradual loss of distinction between the two 
types of caustics.  The condition for the merging of the two caustics is 
that the separation between the two caustics is smaller than the half of 
the sum of the individual caustic widths, i.e.\
\begin{equation}
{\Delta\xi_{c,{\rm c}}+\Delta\xi_{c,{\rm p}}\over 2} \geq 
\left\vert s-{1\over s} \right\vert.
\label{eq29}
\end{equation}
By using the analytic expressions for $\Delta\xi_{c,{\rm p}}$
(eqs.~[\ref{eq8}] and [\ref{eq18}]) and $\Delta\xi_{c,{\rm c}}$
(eq.~[\ref{eq24}]), we compute the region of the caustic merging 
in the parameter space of $(s,q)$ and presented in Figure~\ref{fig:five}
(the region enclosed by thick dashed lines).  The region is confined in a 
small region around $|\svec|\sim 1$, but the width of the region increases 
as $q$ increases because the caustic size increases with the increase 
of $q$.

\subsection{Validity of the Approximation}

Are the presented analytic expressions based on perturbative approximation 
good enough for the description of the caustics in planetary microlensing?  
We answer this question by comparing the two sets of caustics constructed 
based on analytic and numerical computations.

In Figure~\ref{fig:six}, we present some pairs of the planetary caustics 
with different values of the planet parameters $s$ and $q$.  In each panel 
of the figure, the blue caustic is drawn by using the analytic expressions 
while the red caustic is the exact one based on numerical computations.
For reference, we note that the mass ratios of the planets with masses 
equivalent to the Jupiter, Saturn, Neptune, and Earth around a host star 
with $\sim 0.3\ M_\odot$ of the most probable Galactic lensing event are 
$q\sim 3\times 10^{-3}$, $10^{-3}$, $2\times 10^{-5}$, and $10^{-5}$, 
respectively.  From the figure, we find that although the deviation increases 
with the increase of the planet/star mass ratio, the analytic approximation 
well describes the planetary caustic in most mass regime of planets 
($q\lesssim [{\cal O}]10^{-3}$).  For the Earth-mass planet, we find that 
the two caustics are eventually indistinguishable.

\section{Conclusion}

We derived analytic expressions for the location, size, and shape of the 
planetary caustic as a function of the star-planet separation and the 
planet/star mass ratio under perturbative approximation.  Based on these 
expressions, we conducted comprehensive analysis on the properties of the 
planetary caustics.  Combined with the analogous expressions for the central 
caustics derived in paper I, we compared the similarities and differences 
between the planetary and central caustics.  We also presented the expressions 
for the size ratio between the two types of caustics and for the condition 
of the merging of the two types of caustics.  These analytic expressions 
will be useful in understanding the dependence of the planetary lensing 
behavior on the planet parameters and thus in interpreting the planetary 
lensing signals.

\acknowledgments 
We would like to thank J.\ H. An and A. Gould for making helpful comments.  
Work by C.H. was supported by the Astrophysical Research Center for the 
Structure and Evolution of the Cosmos (ARCSEC) of Korea Science and 
Engineering Foundation (KOSEF) through Science Research Center (SRC) 
program.

\end{document}